\documentclass[12pt]{article}
\usepackage{graphicx}
\begin{document}
 \noindent {\footnotesize\it Astronomy Letters, 2010 Vol. 36, No. 3}

 \noindent
 \begin{tabular}{llllllllllllllllllllllllllllllllllllllllllllll}
 & & & & & & & & & & & & & & & & & & & & & & & & & & & & & & & & & & & & &  \\\hline\hline
 \end{tabular}

 \vskip 1.5cm
 \centerline {\large\bf Searching for Stars Closely Encountering with the Solar System}
 \bigskip
 \centerline {V.V. Bobylev}
 \bigskip
 \centerline {\small\it
 Pulkovo Astronomical Observatory, Russian Academy of Sciences, St-Petersburg}
 \bigskip

{\bf Abstract}--Based on a new version of the Hipparcos catalog
and currently available radial velocity data, we have searched for
stars that either have encountered or will encounter the solar
neighborhood within less than 3 pc in the time interval from $-2$
Myr to $+2$ Myr. Nine new candidates within 30 pc of the Sun have
been found. To construct the stellar orbits relative to the solar
orbit, we have used the epicyclic approximation. We show that,
given the errors in the observational data, the probability that
the well-known star HIP 89 825 (GL 710) encountering with the Sun
most closely falls into the Oort cloud is 0.86 in the time
interval $1.45\pm0.06$ Myr. This star also has a nonzero
probability, $1\times10^{-4},$ of falling into the region $d<1000$
AU, where its influence on Kuiper Belt objects becomes possible.

\bigskip

DOI:10.1134/S1063773710030060

\section*{INTRODUCTION}

Interest in the problem of close encounters of field stars with
the Solar system stems from the fact that the passage of a star
can lead to various kinds of perturbations of Solar system
objects.

Thus, for example, the interaction of a star with the Oort comet
cloud can give rise to comet showers reaching the region of the
major planets (Hills 1981; Weissman 1996). The formation time
scale of a comet shower is $\approx$1 Myr after the passage of a
star. According to Weissman (1996), the Oort cloud is a spheroid
with a semimajor axis of about $10^5$ AU ($\approx$0.5 pc, since
1~pc=206 265 AU) elongated toward the Galactic center and a
semiminor axis of about $8\times10^4$ AU. The question about close
encounters of stars with the Sun within $r<2-5$ pc was considered
by Revina (1988), Matthews (1994), and Mullari and Orlov (1996)
using various ground-based observations and by Garcia-Sanchez et
al. (1999, 2001) based on Hipparcos (1997) data in combination
with stellar radial velocity data. As a result, 156 Hipparcos
stars from the solar neighborhood 50 pc in radius that either have
encountered or will encounter with the Solar system within $r<5$
pc in the time interval $\pm10$ Myr are known to date
(Garcia-Sanchez et al. 2001). Having analyzed these data,
Garcia-Sanchez et al. (2001) estimated the frequency of close
(within 1 pc) encounters of stars with the Sun to be $2.3\pm0.2$
encounters per Myr and, after a correction for the Hipparcos
incompleteness, this value increases to $11.7\pm1.3$ encounters
per Myr.

Analyzing the possibility of even closer encounters is of current
interest. For example, Kenyon and Bromley (2004) argue that only
the passage of a star can explain the peculiarities of the orbit
of the minor planet 2003 VB12 (Sedna). The various kinds of
influences of such a star on Kuiper Belt objects were shown to
manifest themselves at characteristic distances $r<1000$ AU
($\approx$0.005 pc). The Kuiper Belt proper extends from the Sun
to a boundary of $\approx$50 AU.

The goal of this study is to search for candidate stars closely
encountering with the Sun based on a new version of the Hipparcos
catalog (van Leeuwen, 2007) and currently available stellar radial
velocity data. We solve the problem of statistical simulations by
taking into account the random errors in the input data and
estimate the probability of a star penetrating into the Oort cloud
region and into the region of a possible influence of the star on
Kuiper Belt objects.

\section{THE DATA}

We use stars from the Hipparcos catalog (ESA 1997) while taking
the new proper motions and parallaxes from a revised version of
this catalog (van Leeuwen, 2007); the stellar radial velocities
are taken from the Pulkovo Compilation of Radial Velocities (PCRV)
(Gontcharov 2006) created as a result of implementing the OSACA
project (Bobylev et al., 2006) and containing radial velocity data
for about 35 000 stars. Note that, in contrast to the larger
CRVAD-2 catalog of radial velocities (Kharchenko et al., 2007),
the PCRV catalog contains only the stars with random errors in
their radial velocities within 10 km s$^{-1}$.

\section{THE METHODS}
\subsection{Orbit Construction}

We use a rectangular Galactic coordinate system with the axes
directed away from the observer toward the Galactic center
($l=0^\circ, b=0^\circ,$ the $X$ axis), in the direction of
Galactic rotation ($l=90^\circ, b=0^\circ,$ the $Y$ axis), and
toward the North Pole ($b=90^\circ,$ the $Z$ axis). The
corresponding space velocity components of an object $U,V,W$ are
also directed along the $X,Y,Z$ axes. The epicyclic approximation
(Lindblad, 1927) allows the stellar orbits to be constructed in a
coordinate system rotating around the Galactic center. We apply
this method in the form given in Fuchs et al. (2006):
$$
\displaylines{\hfill
 X(t)= X(0)+{U(0)\over \kappa} \sin(\kappa t)+     \hfill\cr\hfill
      +{V(0)\over 2B} (1-\cos(\kappa t)),         \hfill\llap(1)\cr\hfill
 Y(t)= Y(0)+2A \biggl( X(0)+{V(0)\over 2B}\biggr) t-\hfill\cr\hfill
       -{\Omega_0\over B\kappa} V(0) \sin(\kappa t)
       +{2\Omega_0\over \kappa^2} U(0) (1-\cos(\kappa t)),\hfill\cr\hfill
 Z(t)= {W(0)\over \nu} \sin(\nu t) + Z(0) \cos(\nu t), \hfill
 }
$$
where $t$ is the time in Myr (we proceed from the ratio pc/Myr =
0.978 km s$^{-1}$); $A$ and $B$ are the Oort constants;
$\kappa=\sqrt{-4\Omega_0 B}$ is the epicyclic frequency;
$\Omega_0$ is the angular velocity of Galactic rotation for the
local standard of rest,  $\Omega_0=A-B$; $\nu=\sqrt{4\pi G
\rho_0}$ is the vertical oscillation frequency, where $G$ is the
gravitational constant and $\rho_0$ is the star density in the
solar neighborhood.

The parameters $X(0),Y(0),Z(0)$ and $U(0),V(0),W(0)$ in the system
of equations (1) denote the current stellar positions and
velocities. The displacement of the Sun from the Galactic plane is
taken to be $Z(0)=17$~pc (Joshi 2007). We calculate the velocities
$U,V,W$ relative to the local standard of rest (LSR) with
$(U,V,W)_{LSR}=(10.00,5.25,7.17)$ km s$^{-1}$ (Dehnen and Binney
1998).

We took $\rho_0=0.1M_\odot$/pc$^3$ (Holmberg and Flinn 2004),
which gives $\nu=74$ km s$^{-1}$ kpc$^{-1}$. We used the following
Oort constants found by Bobylev et al. (2008): $A=15.5\pm0.3$ km
s$^{-1}$ kpc$^{-1}$ and $B=-12.2\pm0.7$ km s$^{-1}$ kpc$^{-1}$;
$\kappa=37$ km s$^{-1}$ kpc$^{-1}$ corresponds to these values.

Note that we neglect the gravitational interaction between the
star and the Sun.

As was shown by Mullari and Orlov (1996), even the simplest linear
approximation provides a sufficient accuracy in the time interval
from $-2$ Myr to $+2$ Myr compared to the method of constructing
the relative orbits (the star relative to the Sun) in the Galactic
potential. We choose the epicyclic approximation, because we are
planning to perform statistical simulations with the computation
of hundreds of thousands of orbits, while the application of more
complex methods requires a huge computational time.

\subsection{Statistical Simulations}

In accordance with the method of Monte Carlo statistical
simulations, we compute a set of orbits for each object by taking
into account the random errors in the input data. For each star,
we compute the encounter parameter, the shortest distance between
the stellar and solar orbits
 $d_{min}=\sqrt{\Delta X^2(t)+\Delta Y^2(t)+\Delta Z^2(t)}.$
The stellar parameters are assumed to be distributed normally with
a dispersion $\sigma$. We add the errors to the equatorial
coordinates, proper motion components, parallax, and radial
velocity of the star using the nominal errors.

\section{RESULTS}

First, we considered the solar neighborhood 30 pc in radius using
$\approx$35 000 stars with known space velocities. For each star,
we constructed the orbit relative to the Sun in the time interval
from $-2$ Myr to $+2$ Myr. In addition to the stars from the list
by Garcia-Sanchez et al. (2001), we found several other Hipparcos
stars. Data on the nine stars found with an encounter parameter
$d<3$ pc are presented in Table 1; Fig. 1a shows the trajectories
for six of these stars.

Figure 1b shows the model trajectories of four stars computed by
taking into account the random errors in the observational data;
300 realizations were obtained for each of the three stars
indicated in the figure.

Table 2 gives characteristics of the stars common to the list by
Garcia-Sanchez et al. (2001) that have the closest ($d<2$ pc)
encounters with the Sun.

Table 3 gives characteristics of the common list of stars that
have the closest ($d<2$ pc) encounters with the Sun. Note that
this is currently the most complete list with the restrictions
that we used. In contrast to Tables 1 and 2, Table 3 gives the
encounter parameters $d_{min}$ and $t_{min}$ with their errors
estimated through statistical simulations.

Table 3 also contains two stars from Table 1. The first of them is
the well-known white dwarf WD 0310.688 (HIP 14754). Note that the
error in its radial velocity (Pauli et al. 2006) is largest among
all of the stars listed in Tables 1 and 2. The data for the second
star, HIP 27887, are very reliable. It is contained in the catalog
of F and G dwarfs of the Geneva-Copenhagen survey (Nordstrom et
al. 2004; Holmberg et al. 2007).

Note that the compilations by Gontcharov (2006) and Kharchenko et
al. (2007) give an obsolete radial velocity for HIP 3829,
$V_r=263\pm5$ km s$^{-1}$, taken from the GCRV (Wilson 1953). A
very close encounter with the Sun is obtained with this value:
$d_{min}=0.96$ pc at $t_{min}=-16$ thousand years. However, this
star is the well-known white dwarf WD~0046+051 (van Maanen’s star,
vMa 2). According to more recent observations, its radial velocity
is $V_r=54$ km s$^{-1}$ (Greenstein and Trimble 1967) or
$V_r(LSR)=-41$ km s$^{-1}$ (Aannestad et al. 1993). In this case,
the encounter is considerably less close, $d_{min}>3$ pc (see also
Garcia-Sanchez et al. 1999, 2001).

Figure 2 shows the trajectories of HIP 89825 (GL 710) relative to
the Sun computed by taking into account the random errors in the
observational data.

We made only 300 realizations for a clearer graphical
presentation. It is obvious from the figure that very close
encounters with the Solar system are possible for this star. More
detailed simulations show that, for example, out of the one
million model orbits found for GL~710, $d_{min}\leq0.48$ pc in
855~902 cases and, hence, the probability of its falling into the
Oort cloud is $P_1=0.86$ and $d_{min}\leq0.005$ pc in 94 cases
and, hence, the probability of the star approaching the boundary
of its possible influence on Kuiper Belt objects is
$P_2=0.94\times10^{-4}$.

Figure 3 show the distribution of encounter parameters of HIP
89825 (GL~710) with the Sun obtained using one million model
orbits.

\section{DISCUSSION}

(1) Among the Hipparcos stars we consider, there are nine stars
common to the list by Mullari and Orlov (1996). The encounter
parameters for some of the stars found by Mullari and Orlov (1996)
are:  $d_{min}=1.35$ pc,  $t_{min}=1050$ thousand years for
GL~710; $d_{min}=1.64$ pc, $t_{min}=-529$ thousand years for
GL~208; and $d_{min}=1.89$ pc, $t_{min}=88.7$ thousand years for
GL~860 A. The encounter parameters agree well with our results
(Table2). There is only one exception, GL~710, for which the new
data give a closer encounter, which was also pointed out by
Garcia-Sanchez et al. (1999).

Note that the star HIP 99461 (GL 783) we selected is present in
the list by Mullari and Orlov (1996) but is absent in the lists by
Garcia-Sanchez et al. (1999, 2001).

(2) For an overwhelming majority of the stars, the encounter
parameters derived here agree well with those from Garcia-Sanchez
et al. (2001).

The differences are most noticeable for the star GL 217.1, more
specifically, $|\Delta d_{min}|=0.38$ pc and $|\Delta
t_{min}|=182$ thousand years. According to Garcia-Sanchez et al.
(2001), who adopted the radial velocity $V_r=20.0\pm3.7$ km
s$^{-1}$ that differs markedly from our value (Table 2), the
encounter parameters for GL 217.1 are: $d_{min}=1.65\pm0.27$ pc
and $t_{min}=-(1046\pm163)$ thousand years.

Computing the model epicyclic orbits for the star GL 217.1 using
the initial data from Garcia-Sanchez et al. (2001) with the
adopted radial velocity error $e_{V_r}=2$ km s$^{-1}$ yields
$d_{min}=1.61\pm0.20$ pc and $t_{min}=-(1066\pm116)$ thousand
years, and these parameters almost coincide with those from
Garcia-Sanchez et al. (2001).

This leads us to conclude that the existing discrepancy is
undoubtedly related to the difference in input data.

As we see from Table 3, GL 217.1 is the most massive,
$M=2M_\odot$, star among our stars. Its improved encounter
parameters, $d_{min}=1.28\pm0.06$ pc and $t_{min}=-(861\pm40)$
thousand years, make it attractive for studying the close passages
that could take place in the past.

(3) The star GL 710 is of greatest interest to us, because it has
a record close encounter with the Sun among all of the candidates
known to date. According to Garcia-Sanchez et al. (2001), the
encounter parameters of GL~710 are: $d_{min}=0.34\pm0.18$ pc and
$t_{min}=(1358\pm41)$ thousand years, in good agreement with those
we found (Table 3).

GL 710 is the only star with a significant difference between the
data of Tables 2 and 3. This suggests that the conditions for its
encounter with the Sun make the stellar orbit very sensitive to
small changes in such initial data as the parallax and radial
velocity. We clearly see from Fig. 3a that the distribution of
minimum encounter distance $d_{min}$ differs from a Gaussian one,
the mode of this distribution is $d_{min}=0.27$ pc, which is lower
than the median value of $d_{min}$ given in Table 3, while the
distribution of encounter times is nearly Gaussian (Fig. 3b).

(4) Our simulations show that among the candidates considered,
only the star GL 710 has a high probability of penetrating into
the Oort cloud region.

\section{CONCLUSIONS}

Based on currently available space velocity data for about 35 000
Hipparcos stars, we searched for stars closely encountering with
the Solar system. For this purpose, we took stars within 30 pc of
the Sun and determined their orbits relative to the Sun based of
the epicyclic approximation in the time interval from $-2$ Myr to
$+2$ Myr. We found nine new candidates in addition to the
well-known list of such stars (Garcia-Sanchez et al. 2001). The
use of the PCRV (Gontcharov 2006) and improved Hipparcos stellar
parallaxes (Leeuwen 2007) made this possible.

As a result, we compiled the currently most complete (given the
restrictions) list of Hipparcos stars that have close ($d<2$ pc)
encounters with the Sun.

For the star GL 217.1, a well-known candidate for a passage close
to the Sun, the new observational data were shown to change
noticeably its previously known encounter parameters with the Sun.
The encounter parameters found here are: $d_{min}=1.28\pm0.06$~pc
and $t_{min}=-(861\pm40)$ thousand years. Improving the radial
velocity for the white dwarf WD 0310--688 (HIP 14754) whose orbit
passed at a distance $d_{min}=1.61\pm0.19$ pc from the solar orbit
about 300 thousand years ago is of current interest. Our
statistical simulations showed that the star GL 710 has not only a
high probability of penetrating into the Oort cloud, $P_1=0.86,$
but also a nonzero probability, $P_2=1\times10^{-4},$ of
penetrating into the region where the influence of the passing
star on Kuiper Belt objects is significant.

 \bigskip
{\bf ACKNOWLEDGMENTS}
 \bigskip

I wish to thank Prof. V.V. Orlov for a careful reading of the
manuscript and his remarks and A.T. Bajkova for a discussion of
the results. The SIMBAD searchable database was very helpful in
the work. This study was supported by the Russian Foundation for
Basic Research (project no. 08--02--00400) and in part by the
``Origin and Evolution of Stars and Galaxies'' Program of the
Presidium of the Russian Academy of Sciences.

 \bigskip
{\bf REFERENCES}
 \bigskip

{\small

Aannestad P.A., S.J. Kenyon, G.L. Hammon, et al., Astron. J. 105,
1033 (1993).

Anosova J., V.V. Orlov, and N.A. Pavlova, Astron. Astrophys. 292,
115 (1994).

Bobylev V.V., A.T. Bajkova, and A.S. Stepanishchev, Astron. Lett.
34, 515 (2008).

Bobylev V.V., G.A. Gontcharov, and A.T. Bajkova, Astron. Rep. 50,
733 (2006).

Dehnen W., and J.J. Binney, MNRAS 298, 387 (1998).

Fuchs B., D. Breitschwerdt, M.A. Avilez, et al., MNRAS 373, 993
(2006).

Garcia-Sanchez J., R.A. Preston, D.L.Jones, et al., Astron. J.
117, 1042 (1999).

Garcia-Sanchez J., P.R.Weissman, R.A.Preston, et al., Astron.
Astrophys. 379, 634 (2001).

Giampapa M.S., R. Rosner, V. Kashyap, et al., Astrophys. J. 463,
707 (1996).

Gontcharov G.A., Astron. Lett. 32, 759 (2006).

Greenstein J.L., and V.L. Trimble, Astrophys. J. 149, 283 (1967).

Hills J.G., Astron. J. 86, 1730 (1981).

The Hipparcos and Tycho Catalogues, ESA SP-1200 (1997).

Holmberg J., and C. Flinn, MNRAS 352, 440 (2004).

Holmberg J., B. Nordstrom, and J. Andersen, Astron. Astrophys.
475, 519 (2007).

Joshi Y.C., MNRAS 378, 768 (2007).

Kenyon S.J., and B. Bromley, Nature 432, 598 (2004).

Kharchenko N.V., R.-D. Scholz, A.E. Piskunov, et al., Astron.
Nachr. 328, (2007).

van Leeuwen F., Astron. Astrophys. 474, 653 (2007).

Lindblad B., Arkiv Mat., Astron., Fysik A 20, (17), (1927).

Matthews R.A.J., R. Astron. Soc. Quart. J. 35,1 (1994).

Mullari A.A., and V.V. Orlov, Earth, Moon, and Planets (Kluwer
Netherlands, 1996), Vol. 72, p. 19.

Nordstrom B., M. Mayor, J. Andersen, et al., Astron. Astrophys.
419, 989 (2004).

Pauli E.M., R. Napiwotzki, U. Heber, et al., Astron. Astrophys.
447, 173 (2006).

Revina I.A., Analysis of the Motion of Celestial Bodies and
Estimation of the Accuracy of Their Observations (Latvian Univ.,
Riga, 1988), p. 121.

Weissman P.R., Earth, Moon, and Planets (Kluwer, Netherlands,
1996), Vol. 72, p. 25.

Wilson R.E., General Catalogue of Stellar Radial Velocities
(Carnegie Inst., Washington, DC, 1953), p. 601.

\bigskip
Translated by N. Samus'

}

\newpage
{
\begin{table}[t]                                                
\caption[]{\small\baselineskip=1.0ex\protect
 Hipparcos stars within 30 pc of the Sun found here
}
\begin{center}
\begin{tabular}{|r|l|r|r|r|r|r|r|r|r|r|}\hline

        &      &            &            &                     &                    &   & \\
    HIP & SP   & $d_{min},$ & $t_{min},$ & $\pi\pm e_\pi,$~~~~ & $V_r\pm e_{V_r},$~ & n & $\varepsilon_{V_r},$ \\
        &      &      pc    & $10^3$ yr  &    mas~~~~~~ &    km s$^{-1}$~~        &   &  km s$^{-1}$\\\hline
    1~~ & ~~2  &  ~~3       & 4 ~~~      &             5~~~~~~ &       6~~~~~       & 7 &  8~~ \\\hline
  14754 & DA   & $1.60$ & $ -296$ & $  97.66\pm1.85$ & $  33.8\pm3.2$ & 1 &     \\\hline
  27887 & K3V  & $1.97$ & $ -413$ & $  76.95\pm0.37$ & $  30.7\pm0.3$ & 1 &     \\\hline
 99461* & K2V  & $2.05$ & $ 41.1$ & $ 166.25\pm0.27$ & $-129.0\pm0.2$ & 2 & 0.8 \\\hline
  24186 & M0V  & $2.15$ & $-10.8$ & $ 255.66\pm0.91$ & $ 244.4\pm0.2$ & 3 & 1.1 \\\hline
 105766 & G5   & $2.20$ & $  341$ & $  37.87\pm0.42$ & $ -76.6\pm0.3$ & 2 & 0.2 \\\hline
  30344 & K0V  & $2.30$ & $-2035$ & $  34.10\pm0.62$ & $  14.6\pm0.3$ & 1 &     \\\hline
  26373 & K0V  & $2.33$ & $ -762$ & $  39.82\pm1.36$ & $  32.5\pm0.3$ & 1 &     \\\hline
 104214 & K5V  & $2.79$ & $ 18.6$ & $ 286.82\pm6.78$ & $ -65.9\pm0.1$ & 5 & 0.7 \\\hline
 104217 & K7V  & $2.80$ & $ 19.6$ & $ 285.88\pm0.54$ & $ -64.3\pm0.1$ & 3 & 0.3 \\\hline

\end{tabular}
\end{center}
 {\small
Note. $n$ is the number of catalogs used to calculate the mean
$V_r;$ $\varepsilon_{V_r}$ is the radial velocity error calculated
from the external convergence (Gontcharov 2006); (*) HIP 99461 (GL
783) is present in the list by Mullari and Orlov (1996) but is
absent in the lists by Garcia-Sanchez et al. (1999, 2001).
 }\vskip60mm
\end{table}
 }

\newpage
{
\begin{table}[t]                                                
\caption[]{\small\baselineskip=1.0ex\protect
Data on the known
Hipparcos stars from the solar neighborhood 30 pc in radius
encountering with the solar orbit within $d<2$ pc

}
\begin{center}
\begin{tabular}{|r|l|l|r|r|r|r|r|r|r|r|}\hline
        &                &            &            &                     &                    &   & \\
    HIP & Обозначение    & $d_{min},$ & $t_{min},$ & $ \pi\pm e_\pi,$~~~ & $V_r\pm e_{V_r},$~ & n & $\varepsilon_{V_r},$ \\
        &                &     ~~pc   & $10^3$ yr  &            mas~~~~~ &      km s$^{-1}$~~        &   &  km s$^{-1}$ \\\hline
    1~~ &   ~~~~~2       &      ~~3   & 4 ~~~      &            5~~~~~~~ &       6~~~~        & 7 &  8~~ \\\hline
  89825 & GL 710         & $0.21$ & $1444$ & $  51.12\pm1.63$ & $ -13.8\pm0.3$ &  4 & 8.1 \\\hline
  70890 & Proxima Cen    & $0.89$ & $27.4$ & $ 771.64\pm2.60$ & $ -25.1\pm0.9$ &    &     \\\hline
  71683 & $\alpha$ Cen A/B * & $0.91$ & $28.4$ & $ 754.81\pm4.11$ & $ -24.7\pm0.4$ &  1 &     \\\hline
  57544 & AC+79 3888     & $1.06$ & $46.0$ & $ 186.86\pm1.70$ & $-111.6\pm0.2$ &  5 & 3.9 \\\hline
  87937 & Barnard's star & $1.15$ & $ 9.8$ & $ 548.31\pm1.51$ & $-110.6\pm0.2$ &  5 & 2.2 \\\hline
  27288 & GL 217.1       & $1.27$ & $-862$ & $  46.28\pm0.16$ & $  24.7\pm1.1$ &  2 & 5.9 \\\hline
  54035 & Lalande 21185  & $1.43$ & $20.5$ & $ 392.64\pm0.67$ & $ -85.8\pm0.1$ &  7 & 3.1 \\\hline
  26335 & GL 208         & $1.54$ & $-500$ & $  88.97\pm1.02$ & $  22.0\pm0.2$ &  5 & 2.1 \\\hline
  57548 & Ross 128       & $1.92$ & $72.4$ & $ 298.04\pm2.30$ & $ -31.0\pm0.2$ &  4 & 1.1 \\\hline
 110893 & GL 860 A       & $1.93$ & $90.0$ & $ 249.94\pm1.87$ & $ -34.0\pm0.1$ &  6 & 2.3 \\\hline
  92403 & Ross 154       & $1.96$ & $ 157$ & $ 336.72\pm2.03$ & $ -10.7\pm0.2$ &  5 & 1.9 \\\hline
  38228 & HD 63433       & $2.02$ & $1366$ & $  45.45\pm0.53$ & $ -16.1\pm0.1$ &  6 & 1.2 \\\hline
\end{tabular}
\end{center}
 {\small
Note. $n$ is the number of catalogs used to calculate the mean
$V_r;$ $\varepsilon_{V_r}$ is the radial velocity error calculated
from the external convergence (Gontcharov 2006); (*) the encounter
parameters were calculated from component A.

 }\vskip40mm
\end{table}
 }

\newpage
{
\begin{table}[t]                                                
\caption[]{\small\baselineskip=1.0ex\protect
 Hipparcos stars from the solar neighborhood 30 pc in
radius encountering with the solar orbit within $d<2$ pc

}
\begin{center}
\begin{tabular}{|r|l|c|c|c|c|r|c|c|c|c|}\hline
        &                &        &           &     &            &    \\
    HIP & Обозначение    & SP     &$M/M_\odot$& Ref & ${\overline d}_{min},$ $10^3$ yr & ${\overline t}_{min},$ pc \\
    1~~ &   ~~~~~~2      &  3     & 4         &  5  &   6        & 7  \\\hline

  89825 & GL 710         & K7V    & $0.4-0.6$              & (1) & $0.311\pm0.167$ & $ 1447\pm60$ \\\hline
  70890 & Proxima Cen    & M5Ve   & $0.15\pm0.02$          & (2) & $0.890\pm0.019$ & $27.4\pm0.1$ \\\hline
  71683 & $\alpha$ Cen A/B * & G2V/K1V & $2.0\pm0.1$         & (2) & $0.910\pm0.012$ & $28.4\pm0.1$ \\\hline
  57544 & AC+79 3888     & M4     & $\approx0.15$          &     & $1.059\pm0.013$ & $46.0\pm0.3$ \\\hline
  87937 & Barnard's star & sdM4   & $0.144$                & (3) & $1.148\pm0.006$ & $ 9.8\pm0.1$ \\\hline
  27288 & GL 217.1       & A2Vann & $2.0$                  & (1) & $1.275\pm0.064$ & $ -861\pm40$ \\\hline
  54035 & Lalande 21185  & M2V    & $0.39$                 & (4) & $1.426\pm0.005$ & $20.5\pm0.1$ \\\hline
  26335 & GL 208         & K7     & $0.47$                 & (4) & $1.537\pm0.050$ & $  -500\pm8$ \\\hline
  14754 & WD 0310-688    & DA     & $\approx0.5$           &     & $1.607\pm0.190$ & $ -297\pm29$ \\\hline
  57548 & Ross 128       & M4.5V  & $0.24$                 & (4) & $1.920\pm0.028$ & $72.8\pm0.7$ \\\hline
 110893 & GL 860 A       & M2V    & $0.56$                 & (4) & $1.929\pm0.028$ & $90.3\pm0.5$ \\\hline
  92403 & Ross 154       & M3.5Ve & $0.23$                 & (4) & $1.959\pm0.030$ & $   157\pm1$ \\\hline
  27887 & HD 40307       & K3V    & $0.75^{+0.78}_{-0.71}$ & (5) & $1.974\pm0.035$ & $  -413\pm5$ \\\hline
  38228 & HD 63433       & G5IV   & $0.88^{+0.98}_{-0.84}$ & (5) & $2.038\pm0.099$ & $ 1366\pm20$ \\\hline
\end{tabular}
\end{center}
 {\small
 Note. The stellar mass estimates were taken from the following
papers: (1) Garcia-Sanchez et al. (1999); Anosova et al. (1994);
(3) Giampapa et al. (1996); (4) Mullari and Orlov (1996); and (5)
Holmberg et al. (2007). If there is no reference, then a typical
mass is given; (*) the sum of the masses is given.
 }
 \vskip40mm
\end{table}
 }

\newpage
\begin{figure}[p]
{
\begin{center}
 \includegraphics[width=120mm]{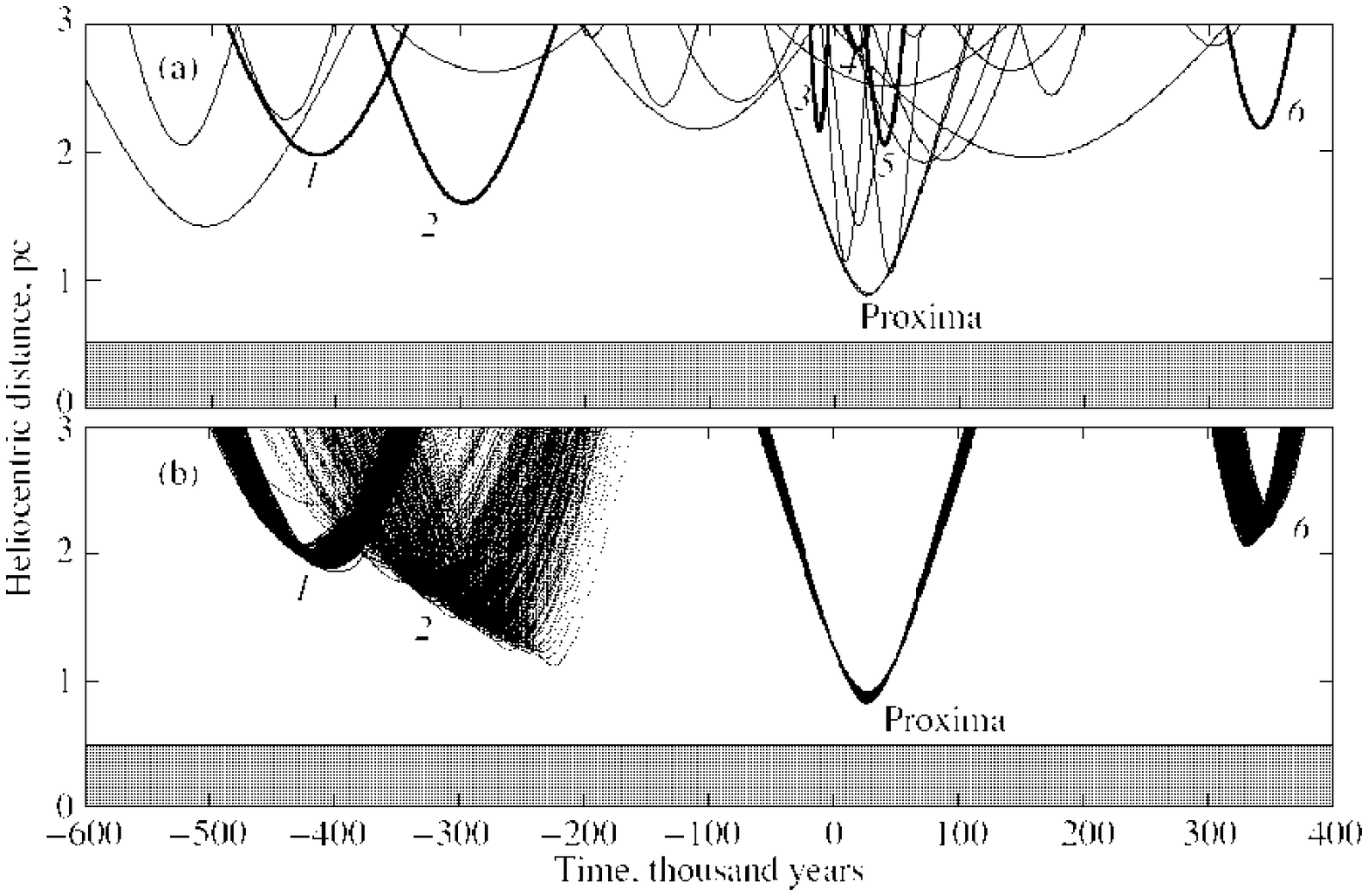}
\end{center}
}
 {\small Fig.~1. (a) Stellar trajectories relative to the Sun. The
trajectories of the stars found here are highlighted by the thick
lines: HIP 27887 (1), HIP 14754 (2), HIP 24186 (3), the pair of
HIP 104214 and HIP 104217 (4), HIP 99461 (5), and HIP 105766 (6).
(b) The model trajectories of four stars computed by taking into
account the random errors in the observational data (300
realizations). The trajectories hatch the $3\sigma$ confidence
regions; the Oort cloud region is shaded.}
\end{figure}

\newpage
\begin{figure}[p]
{
\begin{center}
 \includegraphics[width=80mm]{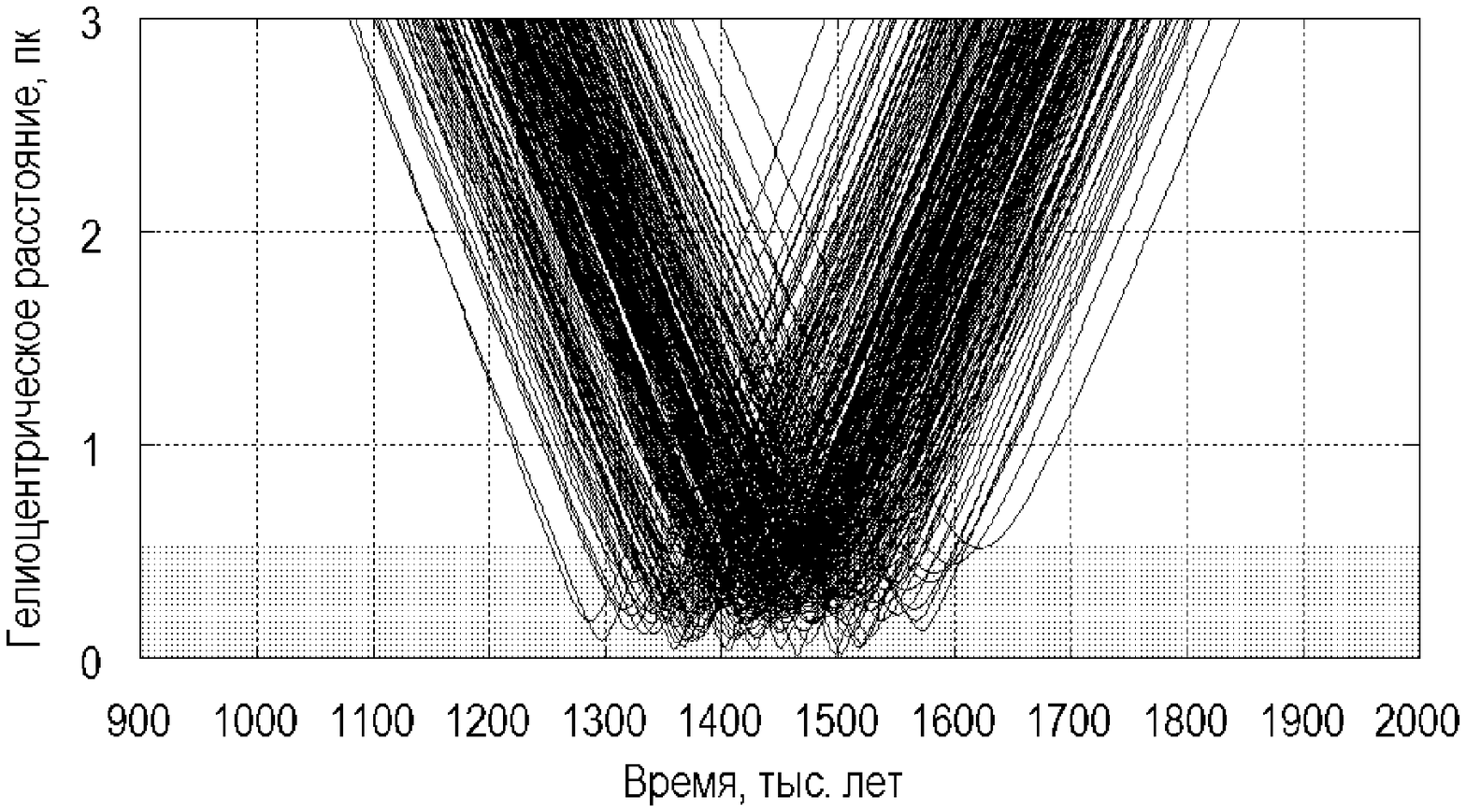}
\end{center}
}
 {\small Fig.~2. Model trajectories of the star HIP 89825 (GL 710) relative
to the Sun computed by taking into account the random errors in
the observational data (300 realizations). The trajectories hatch
the $3\sigma$ confidence region; the Oort cloud region is shaded.}
\end{figure}

\newpage
\begin{figure}[p]
{
\begin{center}
 \includegraphics[width=140mm]{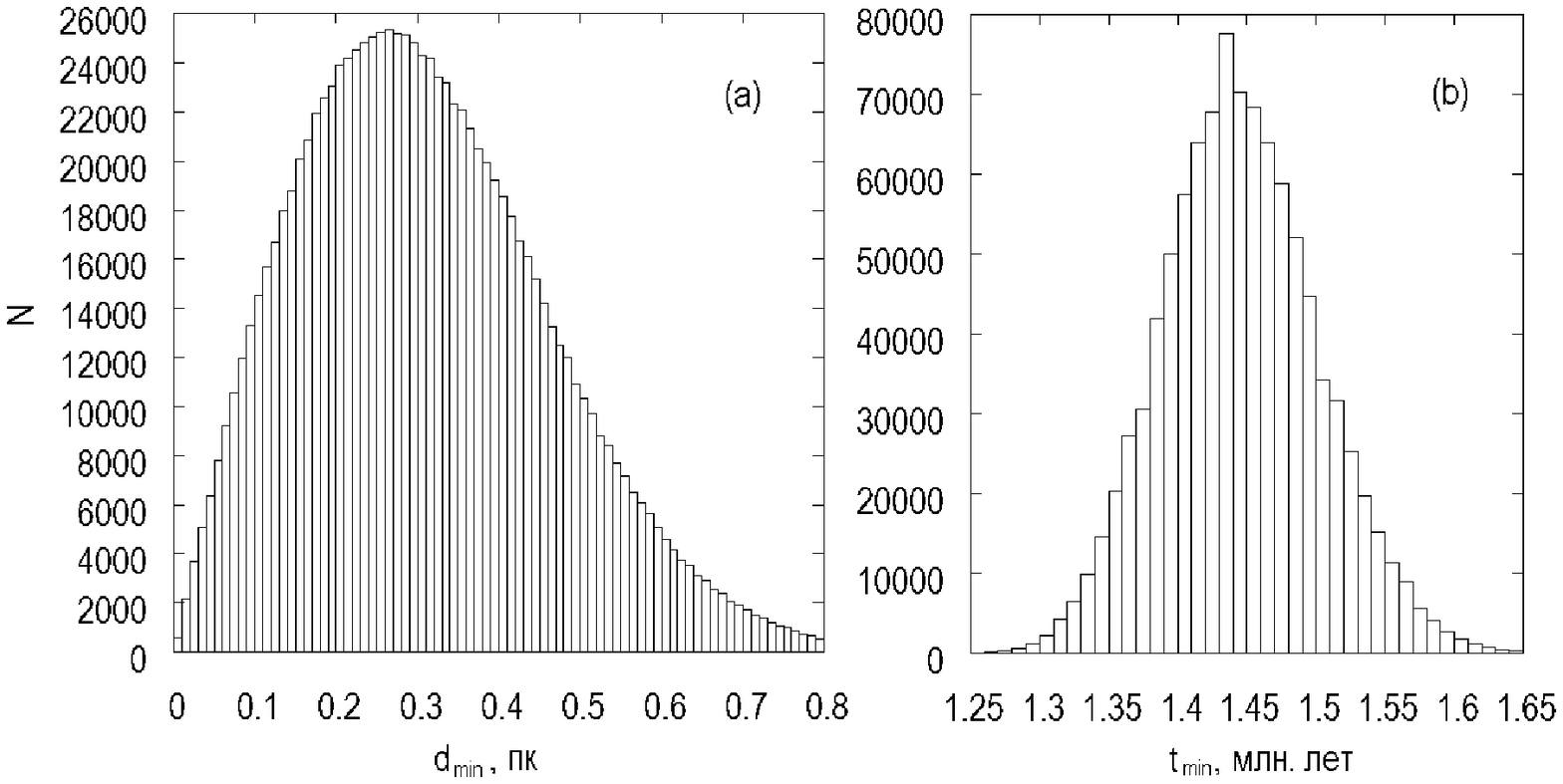}
\end{center}
}
 {\small Fig.~3. Model encounter parameters of the star HIP 89825
(GL 710) with the Sun. (a) The distribution of minimum distance
$d_{min}$; (b) the histogram of encounter times $t_{min}$.}
\end{figure}
\end{document}